# High Accuracy Visible Light Positioning Based on Multi-target Tracking Algorithm

Linyi Huang[1,2,#], Wentao Yang[3,#], Shangsheng Wen[1,*], Manxi Liu[2], Weipeng Guan[2]
1. School of Materials Science and Engineering, South China University of Technology, Guangzhou, 510640, China
2. School of Automation Science and Engineering, South China University of Technology, Guangzhou, 510640, China
3. School of Electronic and Information Engineering, South China University of Technology, Guangzhou, 510640, China

\# These authors contributed equally to this work, Linyi Huang and Wentao Yang are co-first authors of the article.
*Corresponding author: Shangsheng Wen (shshwen@scut.edu.cn)

**Abstract:** In this paper, we propose a multi-target image tracking algorithm based on continuously adaptive mean-shift (Cam-shift) and unscented Kalman filter. By dynamically adjusting the weights of the multi-target motion states, the algorithm achieved better robustness in the case of occlusion, the real-time performance to complete one positioning and relatively high accuracy. The results of the tracking algorithm are evaluated with the tracking error we defined. Then combined with the double-lamp positioning algorithm, the real position of the terminal is calculated and evaluated with the positioning error we defined. Experiments show that the defined tracking error is 0.61cm and the defined positioning error for 3-D positioning is 3.29cm with the average processing time of 91.63ms per frame. Even if nearly half of the LED area is occluded, the tracking error remains at 5.25cm. All of this shows that the proposed visible light positioning (VLP) method can track multiple targets for positioning at the same time with good robustness, real-time performance and accuracy. In addition, the definition and analysis of tracking errors and positioning errors indicates the direction for future efforts to reduce errors.

## 1 Introduction

The Internet of Things(IoT) brings people and devices closer and closer. As a result, more and more scenarios require more accurate and robust indoor positioning systems (IPS) to meet their needs for location based services (LBS) [1-3]. Due to the poor indoor performance of global positioning system (GPS), many indoor positioning technologies based on wireless electromagnetic wave communications have proposed in the past ten years. Among them, the visible light positioning (VLP) has the advantages of no additional signal access equipment, no electromagnetic interference and so on, and has become a very promising indoor positioning method. There are two main approaches in indoor VLP: photodiode-based (PD-based) and image-sensor based (IS-based) [4]. IS-based VLP method is widely studied as its the ability to distinguish between multiple light sources, anti-interference light, the ability to provide position-related information [5] and no need for complicated mechanical structures. Therefore, it has become very popular with the widespread application of complementary metal oxide semiconductor (CMOS) image sensors.

So far, many IS-based VLP methods have been proposed. In [6], three-dimensional positioning at the centimeter level is achieved using two LED lamps and an image sensor.

In [7], a method based on specular reflections cancellation achieved a high accuracy of 1.78cm. In [5], the method of identifying circular lamp was used. The angle of the camera was obtained by fitting the ellipse to achieve a positioning accuracy of 17.52cm. In addition to pure CMOS sensor, a method to get the angle through Inertial Measurement Unit (IMU) and fuse it with image information is proposed in [8]. In actual scenarios, robustness is also a very important yardstick. However, most current VLP method only considered accuracy but ignored robustness. Robustness has always especially been a bottleneck hindering the development of for VLP and even visible light communication (VLC) technology. People walking or obstructions in the indoor environment inevitably cause shadow effects [9] during the detection of region of interest (ROI). Once the LED signal source is completely or partially occluded, it will cause the traditional pixel intensity-based (row and column scanning, circle detection, Hough transform and other edge detection) ROI detection to fail, resulting in positioning failure. Another tricky problem is real-time ability. In order to achieve real-time positioning of the mobile terminal, each image must be captured and processed in real time to obtain the pixel coordinate of ROI of the LED. Then real-time 3D position coordinate calculation is performed through the LED-ID recognition and positioning algorithm. If the traditional ROI region extraction method is used for the image of each frame searching the whole picture for the LED, a complete search is performed for each picture, requiring a large amount of calculation. Therefore, it is necessary to use the multi-target tracking algorithm to detect the ROIs of the LED. Based on the information of the previous ROIs of the LED, the position of the ROIs in subsequent images can be predicted based on current motion state. Starting searching from these predicted ROIs will significantly reduce the search iterations, which greatly improves the real-time ability of positioning for dynamic targets, especially for multiple targets.

In this paper, we propose a multi-target image tracking algorithm using Cam-shift with unscented Kalman filter (UKF) to track multi-LED. Although our previous work [11] based on the idea that dynamically tracking the movement of the LEDs in the video to estimate the optimal position of the ROI by the state of the movement can improve real-time ability achieved the tracking error of 0.86 cm, the optical flow with Bayesian prediction still needs a very large computation cost. Our previous work hints that compared with single-target tracking, the difficulty of multi-target tracking is to accurately detect multiple ROIs while reducing the amount of calculations to ensure the real-time ability of the system. Considering the poor robustness of the optical flow when the LED is occluded, we introduce mean shift (MS) to particle filter in [12] improving the robustness and real-time ability, but the tracking error is 55.2% higher than [11]. [13] proved that introducing the UKF to the MS algorithm can greatly reduce the calculation amount while improving the tracking accuracy. Inspired by [13], we introduce Cam-shift to our multi-target image tracking algorithm for VLP system and improved it to better integrated with the UKF. With UKF predicting the most probable position of ROIs through the current motion states, the proposed improved Cam-shift greatly reduces the search candidate area of ROI regions, thereby significantly enhancing the real-time performance of the algorithm. The robustness of the UKF ensure that even if the LED is partially occluded, the results are still relatively correct. We introduce weights for the accuracy of position measurement in UKF. The higher the reliability of LED position tracking, the smaller the weight, which means that the UKF will trust the targets tracking results more. Through the algorithm's automatic adjustment of the weight, the influence of simply treating state noise and measurement noise as Gaussian white noise in [11] will be reduced. Since the Bhattacharyya distance can measure the similarity between two discrete probability distributions, we choose the Bhattacharyya coefficient to measure the different between the predicted coordinate of the UKF and the position obtained by the Cam-Shift algorithm as reliability of LED position tracking. Besides, tracking error defined in the proposition is the error between the result calculated by tracking algorithm and the real position of targets in the images. Positioning error defined in the proposition is the error between the position calculated by the positioning algorithm and the position of the terminal in the real world. The work we have done previously [11,12,14,15] used tracking error to simulate the theoretical positioning error. Although the theoretical positioning error can reflect the performance of the algorithm, there are other factors in practice. Therefore, it is significant to distinguish the tracking error and the positioning error. In order to improve our work on error analysis, this paper not only analyzed the tracking error of our proposed algorithm, but also analyzed the positioning error between the real 3-D position of the robot and the position calculated by the algorithm, making the experiment more accurate and closer to the actual application. Experiments show that the introduction of multi-target tracking algorithm into VLP system greatly reduces the computation cost, enables the dynamic tracking of the double lamps in real time, avoids the error caused by the single lamp measurement, and increases the robustness of the system. It also points out the main cause of the whole system error, showing the way for future work to reduce errors.

The content of this paper is organized as follows. Section 2 describes the specific steps of the image tracking and positioning algorithm used in our method. The experimental method steps and analysis are presented in Section 3. Finally, we summarized our work in Section 4.

## 2 PRINCIPLE

Our proposed method uses a rolling shutter for a CMOS image sensor. The exposure and data transfer of the rolling shutter is performed in a row. That is, when the line ends, the data of one line is read out. Different LED use different modulation frequencies, allowing the LED-ID algorithm to distinguish between different LED targets by the rolling shutter effect (RSE) [16]. When a light first appears in the image, the LED-ID algorithm introduced in [17] is used to identify the ID of the LED and initialize the Cam-shift and UKF. The predicted position of the UKF is then taken as the starting position of the iteration in the Cam-shift. When the ratio of the area factor of the target to its initial value are less than the threshold by more than 5 frames, meaning that the LED is lost during the previous tracking process, the LED-ID algorithm will work again to search for the initial position of the LED to start the next tracking cycle. After tracking the ROI of the LEDs, a mapping from the pixel coordinates to the world coordinates is established by the double-lamp positioning algorithm to calculate the actual position of the terminal. The flow chart of the whole algorithm is shown in Fig. 1.

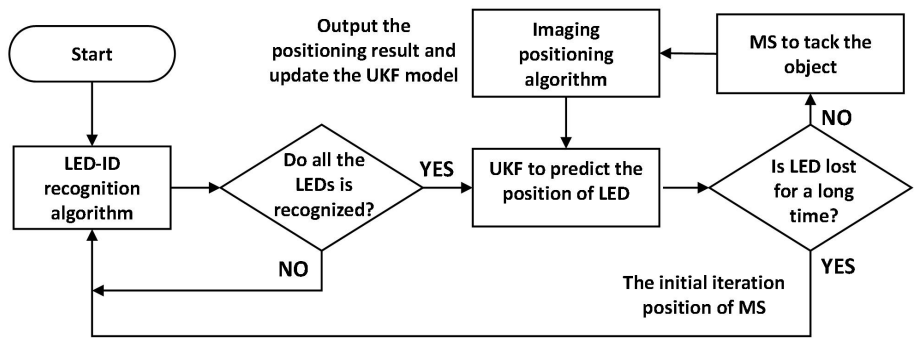

Fig.1. The flow chart of our proposed visible light positioning (VLP) method.

## 2.1 The Proposed Image Tracking Algorithm

Demodulation and positioning require accurate and real-time determination of ROI. In this chapter, we propose a VLP dynamic tracking method based on a multi-target tracking algorithm. By analyzing the video sequence acquired by the image sensor, after detecting the ROI of the LED in the initial image frame, the ROI region of the LED in the subsequent frames is estimated in real time and dynamically according to UKF. More lamps can provide more redundant information to improve the accuracy of positioning, but at the same time reduce real-time performance. Therefore, in order to balance the accuracy and real-time performance, as well as the installation density of lamps in practice, we choose a double-lamp positioning algorithm to process the ROIs obtained by the proposed image tracking algorithm. Based on this idea, this chapter proposes a multi-target tracking algorithm based on improved Cam-shift with unscented Kalman filter to achieve the detection of ROIs.

### 2.1.1 The Proposed Improved Cam-shift

The traditional mean-shift algorithm has been widely studied. Therefore, the basic mathematical method of mean-shift is not repeated in this article. If readers want to know the basic algorithm of Cam-shift, please refer to our previous work [12]. Traditional continuously adaptive mean-shift use the position and size of the search windows in the last frame. Although the position of ROI may be near to the last, the search area is still large if the motion speed of the positioning terminal is high. Therefore, our proposed Cam-shift combines the motion state of the target through unscented Kalman filter. In simple terms, the core of Cam-shift is to find the candidate sample point most similar to the target through non-parametric estimation. To indicate the statistical property of the sample point to the central point, the concept of kernel density function is introduced, giving each sample point a different weight associating with their distance to the center point. Also, the mean shift procedure is guaranteed to make the kernel function converge at a nearby point where the estimate density gradient is zero.

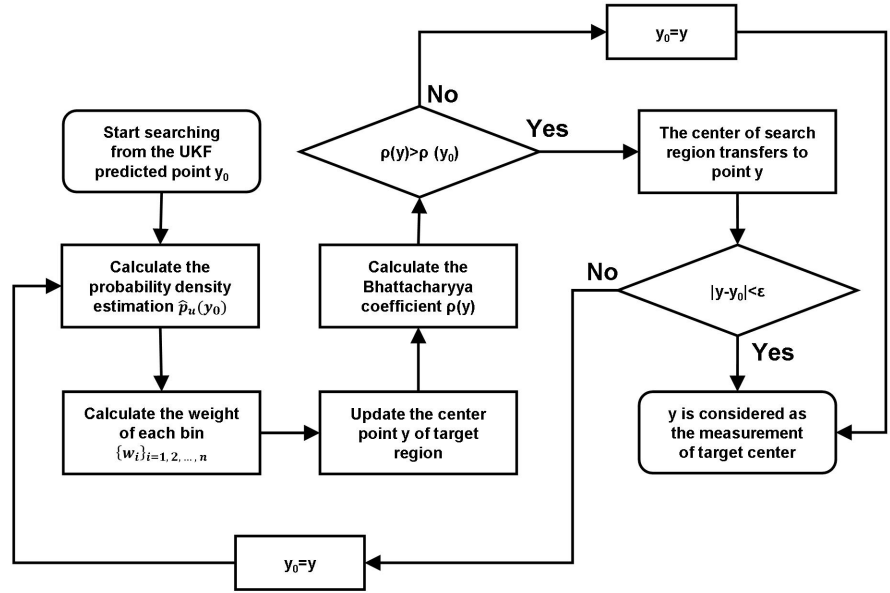

Fig. 2. The process of the proposed Cam-shift algorithm.

It can be seen that whether the Cam-shift algorithm can find the target quickly depends on where the search is started. If the initial position is close to or even the same as the target position, the mean shift has little computational load. Therefore, the key of the improved Cam-shift is to better estimate where to start searching. This is the important reason for the introduction of the UKF.

### 2.1.2 The Unscented Kalman Filter

Although the traditional mean shift algorithm, or traditional Cam-shift algorithm, has good real-time and accuracy under ideal conditions, for an object moving far away from the previous frame at a high speed, the number of iterations of the search is greatly increased. Since the center and size of the search window of the previous frame are used as the initial value of the search for the next frame, this serious influence the real-time performance of the algorithm. Therefore, we use UKF to predict the position of objects in the next frame, which greatly reduces the computational cost of search.

Considering the relative parallelism between the lamp and is, it means that the motion state of different lamps in the image should be similar. Therefore, different from the traditional cam shift, for multi-lamp positioning, we dynamically adjust the motion parameters calculated by different lights according to the occluded area to get a common motion state. Thus, the robustness to occlusion is greatly improved.

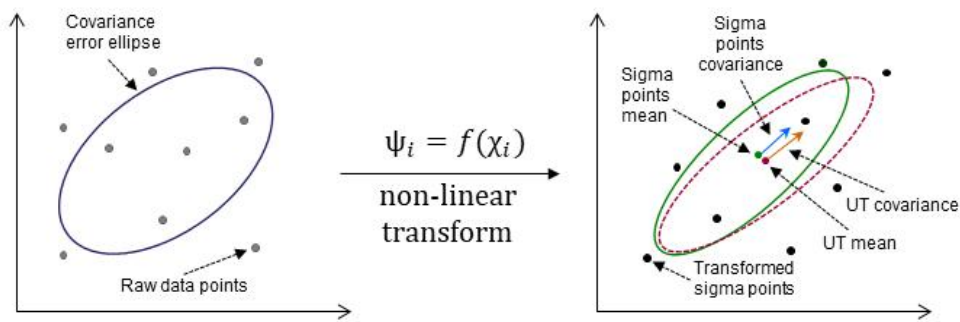

Fig. 3. The schematic diagram of unscented transformation process.

The UKF performs the unscented transform (UT) shown in Fig. 3. on the sampling point, making it suitable for nonlinear systems. By introducing UKF into the Cam-shift algorithm, our proposed algorithm has the ability to track fast moving targets because the mean shift algorithm searches for candidate targets from the UKF predicted position instead of the centroid of the target in last frame. This greatly reduces the number of iterations of the MS. The candidate search area in the current frame is reasonably selected by the UKF, taking into account the a priori position of the target, thus solving the problem that the high-speed motion of the target moves outside the candidate search area within one frame time. At the same time, the real-time performance of the algorithm is also improved due to the reduction in the average number of iterations. When the target is found, its current position is compared to the previous position to estimate its speed and the most likely position in the next frame, through which the UKF state model is updated. The combination of UKF and Cam-shift algorithms enables closed-loop tracking systems to efficiently track fast moving targets in real time.

### 2.2 The Double-lamp Positioning Algorithm

To quantitatively describe the process of image sensor-based VLC positioning, the relevant coordinate system is first defined. The geometrical relationship between the LED and image sensor, shown in Fig. 4., are analyzed to get the location of the positioning terminal in the real world. Notice that the midpoint of lens may not project vertically at the center of image sensor in practice.

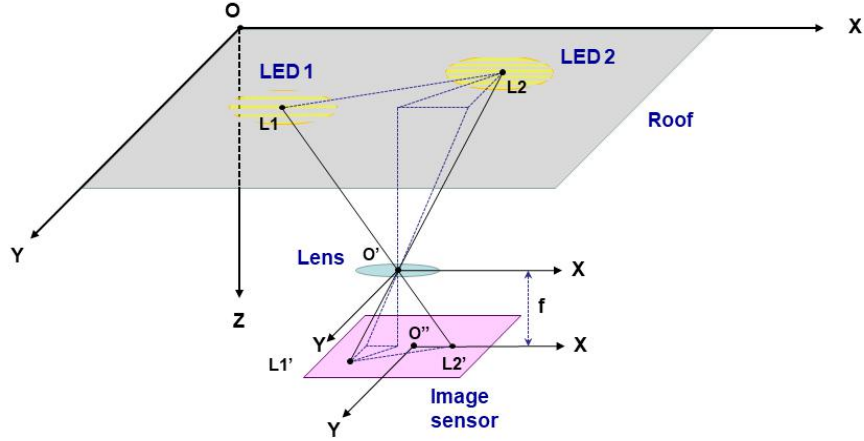

Fig. 4. The geometrical relationship between the LEDs and the image sensor.

Both the pixel coordinate system and the image coordinate system are located on the imaging plane of the image sensor, but the original coordinate points and the measurement unit of the two are inconsistent, as the photosensitive cells are not strictly square. Define $L^t$ as the set of pixels belonging to LED at time t in the image, and $(x_i^t, y_i^t) \in L^t$. i represents ith pixel. The image coordinates is defined as $z_l(t)$ $(x_l^t, y_l^t, z_l^t)$, where dx and dy represent each pixel's size on image sensor in different coordinate directions, respectively. With all the conditions above, the image coordinate $(x_0, y_0)$ of the LED's centroid can be calculated as follow:

$$\begin{bmatrix} x_0 \\ y_0 \\ 1 \end{bmatrix} = \begin{bmatrix} 1/dx & 0 & x^t \\ 0 & 1/dy & y^t \\ 0 & 0 & 1 \end{bmatrix} \begin{bmatrix} x_i^t \\ y_i^t \\ 1 \end{bmatrix} \quad (1)$$

The imaging plane of the image sensor is parallel to the plane of the LED lamp. Therefore, according to the similar triangle, the distance H between the image sensor and the LED lamp can be obtained as follows:

$$H = f \frac{d_{12}}{p_{12}} \quad (2)$$

Where the f is the f is the known camera focal length, $d_{12}$ is the real distance between two LEDs and $p_{12}$ is the distance between two LEDs in image coordinates.

Let the image coordinates and world coordinates of LEDs are (ik,jk) and (xk,yk). For two lamp positioning, k=1,2. When the image coordinate system is parallel to the world coordinate system and the directions of the coordinate axes are the same, the position of terminal can be calculated through a similar triangle relationship:

$$\frac{x - \frac{x_1 + x_2}{2}}{H} = \frac{\frac{i_1 + i_2}{2}}{f} \quad (3)$$

$$\frac{y - \frac{y_1 + y_2}{2}}{H} = \frac{\frac{j_1 + j_2}{2}}{f} \quad (4)$$

## 3 EXPERIMENTS AND DISCUSSION
### 3.1 Experiment Facilities

The experiment facilities include a direct-current voltage source, Turtlebot3 robot (a personal robot kit with robot operating system), an industry camera, 4 LEDs, a personal computer (Hasee Z7M-KP7GT, Windows 10, 8G RAM i7-7700HQ CPU@2.80GHz, Hasee, Shenzhen, China.) and a high-speed video transmission cable. The LEDs powered by the direct-current voltage source were controlled by signal transmitters. The Turtlebot3 robot which can move along a fixed route following a script written in advance is used to carry the industry camera. The long high-speed video transmission cable is used to transfer video data from the industry camera to the PC. The Turtlebot3 with more than one light simulated the moving objects occur in many indoor navigation application scenarios. The PC received the sequential images captured by the industry camera and processed them with our proposed algorithm in real time.

The size of our experiment platform was 190 cm × 100 cm × 190 cm. Three of the LEDs were used to implement the VLP system. The specific parameters of the industry camera, Turtlebot3, direct-current voltage source, and experiment platform are shown in Table 1.

Table 1. Parameters of experiment facilities and platform.

| Camera Specifications | |
|---|---|
| Model | MV-U300 |
| Spectral Response Range/nm | 400~1030 |
| Resolution | 2048 × 1536 |
| Frame Rate/FPS | 46 |
| Dynamic Range/dB | >61 |
| Signal-to-noise Ratio/dB | 43 |
| Pixel Size/μm2 | 3.2 × 3.2 |
| Time of Exposure/ms | 0.0556–683.8 |
| Sensitivity 1.0 V/lux-sec | 550 nm |
| Optical Filter | 650nm Low Pass Optical Filter |
| Type of Shutter | Electronic Rolling Shutter |
| Acquisition Mode | Successive and Soft Trigger |
| Working Temperatures/°C | 0–50 |
| Support Multiple Visual Software | OpenCV |
| Turtlebot3 Robot Specifications | |
| Size (L × W × H)/ mm3 | 138 × 178 × 192 |
| Actuator | Dynamixel XL430-W250 |
| Maximum translational velocity/(m/s) | 0.22 |
| Maximum rotational velocity/(deg/s) | 162.72 |
| Experimental Platform Specifications | |
| Size (L × W × H)/ cm3 | 190 × 100 × 190 |
| LED Specifications | |
| Coordinates (x, y, z)/cm | LED1(100,45,190) LED2(100,145,190) LED3(0,145,190) |
| The half-power angles of LED/deg( $\psi 1/2$) | 60 |

| Circuit Board Specifications | |
| --- | --- |
| Drive chip | DD311 |
| Drive current/A | 0.1 |
| Drive voltage/V | 28 |

Our algorithm is implemented with C++ using the open source computer vision library (Opencv3.4.5). The multi-lamp positioning technique was introduced into our VLP systems, meaning that we need not to acquire all the coordinates of LED but just those we wanted to track, which reduced the complexity of the algorithm and enhanced the real-time ability of our VLP algorithm.

## 3.2 Experiment performance
### 3.2.1 Tracking performance

Positioning accuracy is a vital index measuring the performance of VLP systems. More than one lamps in different shape were introduced into our experiment to test if our algorithm could successfully distinguish the different lamps' ID and obtain their coordinates. In the experiment, two specific lamps with similar shape were chosen to analyze the accuracy of our algorithm. Unlike [5], the improved Cam-shift algorithm can track lamps in different shapes as the result shown in Fig. 5, which can maximize the use of existing luminaries and reduce the cost of system promotion and application. In the figure, lamp 0 is a pair of fluorescent tubes installed on the ceiling as an interference light source without modulation. The LED-ID algorithm can distinguish these lights, and then choose whether to track. The fluorescent tubes are not listed in the experimental facilities as it is not included in the experimental platform. Lamp 1 and 2 are LED1 and LED2 installed on our experimental platform. Images obtained by the industrial camera have been calibrated to eliminate errors caused by optical distortion.

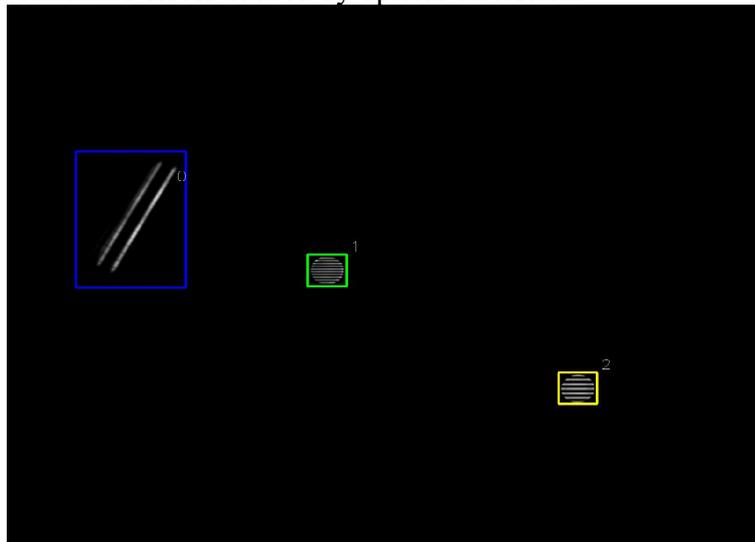

Fig. 5. The tracking performance with multi-lamp in different shapes (0) a pair of fluorescent tubes as interference light source (1) LED1 listed in Table 1 (2) LED2 listed in Table 1

To measure the positioning accuracy of the proposed algorithm in our experiment, 120 continuous video frames were chosen to be analyzed. For comparison, we treat the result got by the pixel intensity detection method used in [17] as the actual position. The result is shown in Fig. 6.

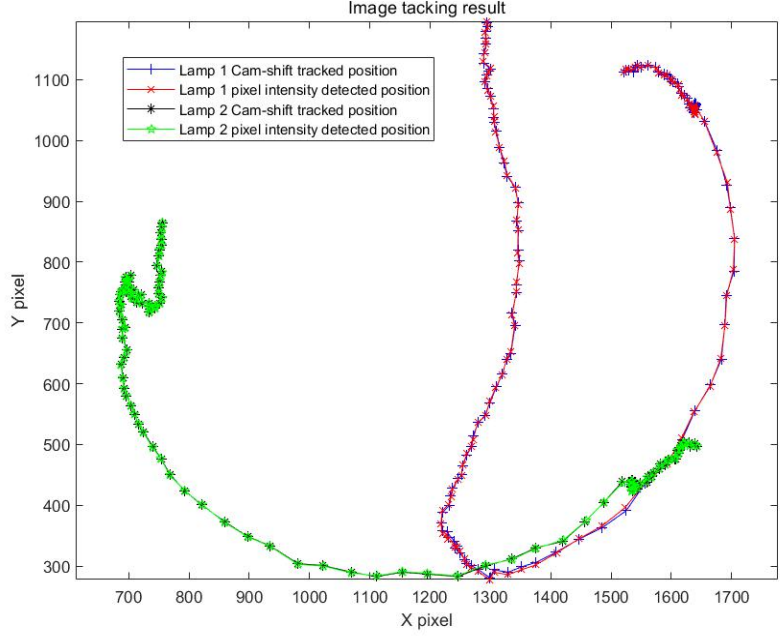

Fig. 6. The tracking result using different algorithm.

From the figure, the trajectory obtained by our proposed Cam-shift tracking algorithm and the pixel intensity detection algorithm almost coincide. In fact, the difference is only 2.66 pixels on average. This means that our proposed tracking algorithm is comparable to the accuracy of pixel intensity detection. Since the pixel error is a relative quantity, it cannot intuitively reflect the influence of tracking error on the overall positioning accuracy. We convert the pixel error into the error of the actual positioning system through our proposed positioning algorithm. For the tracking of a single lamp, since our algorithm requires position data of two lamps, we assume that the data of the other lamp is accurate. It can be concluded that the effect of the pixel tracking error on the double-lamp joint positioning error is an average of 0.379cm.

For more detailed analysis, we define tracking error as:

$$D = \sqrt{(x-x_a)^2 + (y-y_a)^2} \quad (5)$$

Where (x, y) is the position calculated by our algorithm and (xa, ya) is the actual position.

The probability distribution of image tracking error shows the stability of the algorithm. Thus, cumulative distribution function (CDF), the integral of the probability density function, was further analyzed. The CDF of tracking error is shown in Fig. 7. The definition of cumulative distribution function is:

$$F_X(x) = P(X \le x) \quad (6)$$

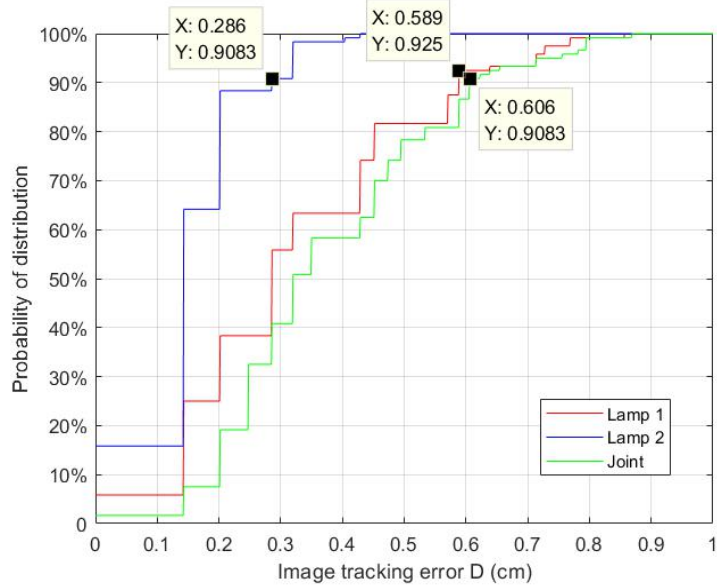

Fig. 7. The cumulative distribution function (CDF) of image tracking error

As the Fig. 7 indicates, more than 90% of tracking error of lamp 1 and lamp 2 are less than 0.286cm and 0.589cm, respectively. Also, more than 90% of tracking error for the two lamps were less than 0.606cm. If we can bear the 10% uncertainty, the tracking error of our algorithm is less than 0.606cm.

### 3.2.2 Positioning performance

In order to better test the performance of our proposed algorithm in practice, we designed the following experiment. The robot moves linearly along the X-axis and Y-axis in a plane range of 80 cm×70 cm to independently indicate the error on the X and Y axes. The height is from the focus of the image sensor to the horizontal plane of the luminaire. Starting from 50cm, there are 5 groups for every 5cm. While the robot is moving continuously, the image is recorded and transmitted to the computer running our proposed algorithm. Turtlebot3 can record the distance traveled and notify the computer to record data when passing certain sampling points. In the end, the result we got is shown in Fig. 8.

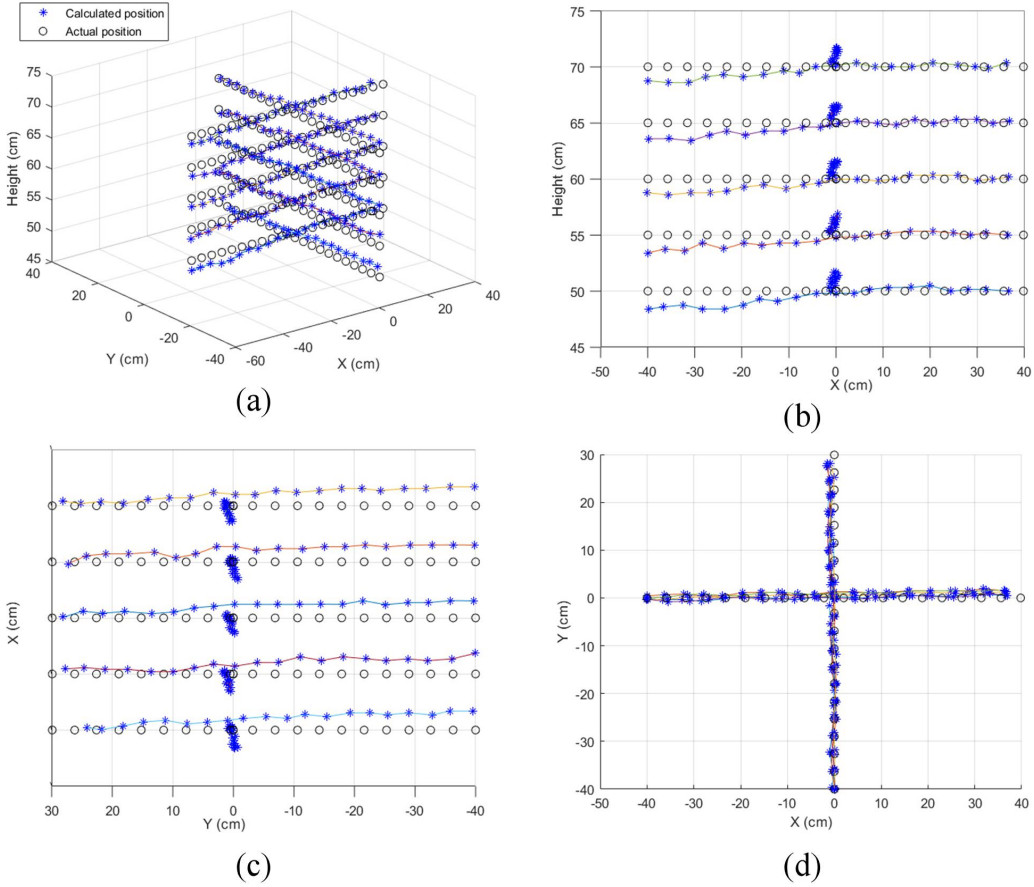

Fig. 8. The tracking position performance. (a) Stereogram. (b) Projecting from y-axis. (c) Projecting from x-axis. (d) Projecting from the z-axis

We apply the image position algorithm to the tracking algorithm we proposed, the pixel intensity detection algorithm used in [17] and the ellipse fitting algorithm used in [5] to compare the performance of different algorithms. We define the three-dimensional positioning error Ds.

$$D_s = \sqrt{(x - x_a)^2 + (y - y_a)^2 + (h - h_a)^2} \qquad (7)$$

Where xa, yb, and ha are the real position coordinates, and x, y, and h are the three-dimensional position calculated by algorithm. The average tracking positioning error of algorithm we proposed, the pixel intensity detection algorithm and the ellipse fitting are 1.97cm, 1.84cm and 1.99cm, respectively. The CDF of different algorithm are calculated and shown in Fig. 9.

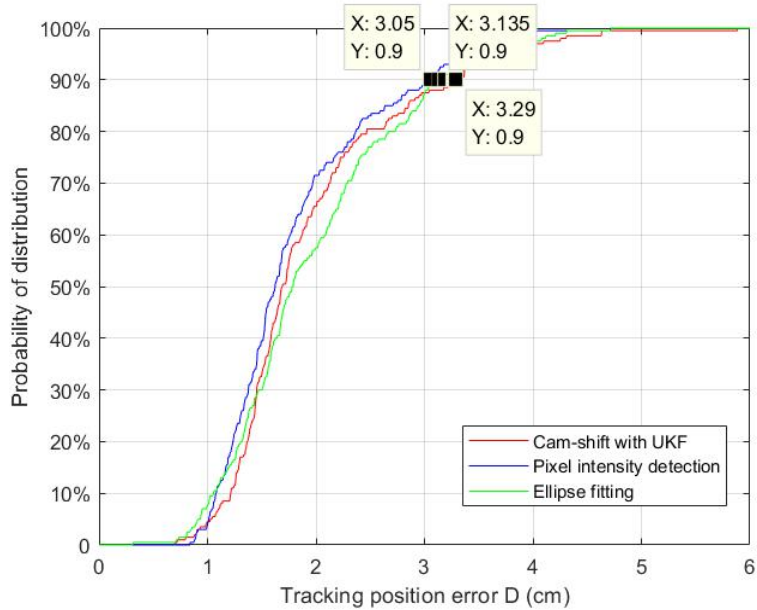

Fig. 9. The CDF of positioning error of different algorithm.

It can be seen that for data more than 90%, the error of algorithm we proposed, the pixel intensity detection algorithm and the ellipse fitting are 3.29cm, 3.05cm and 3.135cm, respectively. It can be seen that three different tracking algorithms perform almost the same under the experimental conditions we have built.

Compare with the tracking error, it can be seen that the error of the entire VLP system mainly comes from the positioning algorithm. Besides, the Mechanical vibration generated by the robot in continuous motion is also one of the sources of error. Therefore, in the future work to improve the positioning accuracy, the focus should be on the more detailed modeling analysis of the position algorithm.

### 3.2.3 Robustness

Robustness is another important feature of VLP systems and was usually overlooked in existing research. In practical applications, the LED may be blocked and the image may be broken. In this case, most VLP methods will fail while our method uses Cam-shift combined with UKF to achieve high positioning accuracy. When the lamp is occluded, the tracking results of the Cam-shift is treated as the observation model of the UKF, combined with the noise matrix to get the final output. In this way, the occluded lamp can be accurately positioned with a small computational cost.

In our experiments, we specifically built tests to evaluate the performance of the algorithm under occlusion conditions. To rule out interference from other factors, we used a black background. The lamps are dynamically and randomly blocked during the experiment with the occlusion area 20% to 90% of the original area. The performance of our proposed algorithm in occlusion area of less than 50% and from 50% to 90% for single lamp and two lamps simultaneously occluded are shown in Fig. 10.

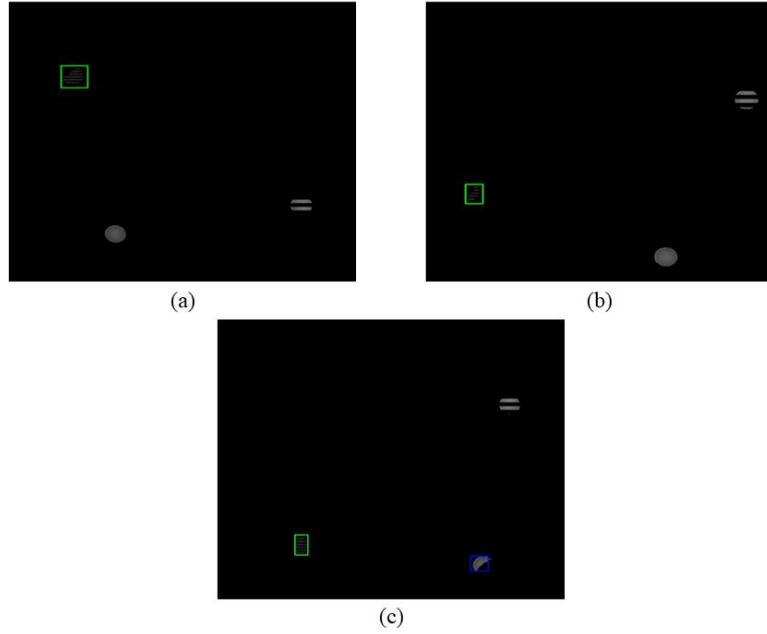

(a)            (b)

(c)

Fig. 10. The tracking performance when two lamps are simultaneously occluded. (a) Single lamp occlusion area less than 50%. (b) Single lamp occlusion area from 50% to 90%. (c) Two lamps simultaneously occluded

For the positioning of a single lamp, when the occlusion area of the LED is between 20% and 50%, the 50 and 40 frames were recorded and analyzed. When the occlusion area is less than 90%, 28 frames were recorded and analyzed. The image tracking error is calculated as shown in Fig. 11.

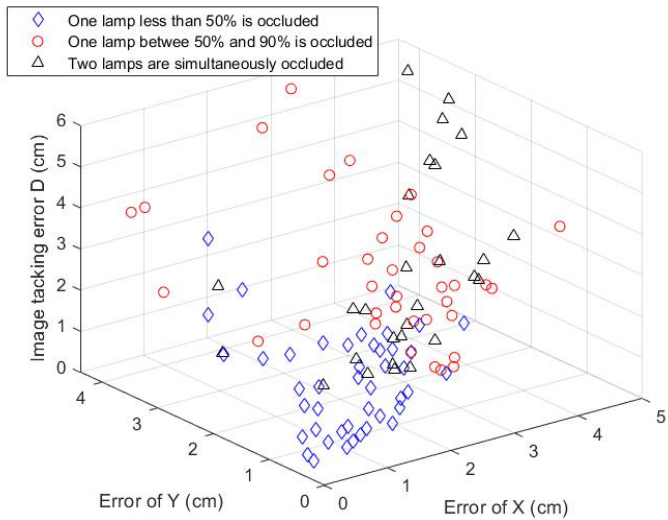

Fig. 11. The image tracking error when lamps are occluded.

When the occlusion area of single lamp is from 20% to 50% and from 50% to 90%, the average tracking error 1.49cm and 3.12cm. When the two lamps simultaneously occluded, the average tracking error is 3.10 cm. Similarly, the CDF of the three different situations are shown in Fig. 12.

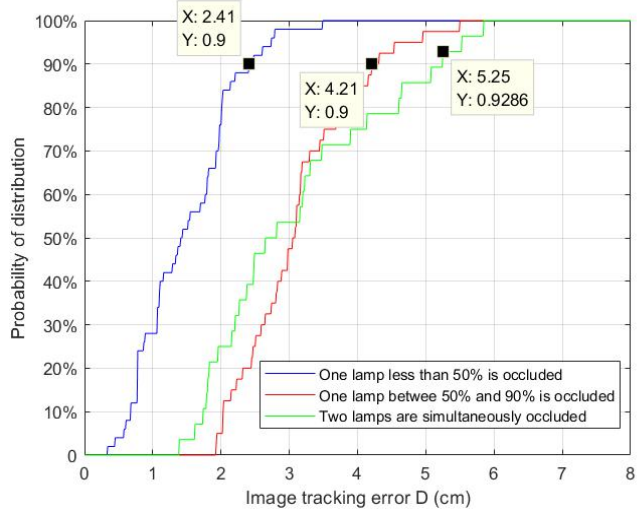

Fig. 12. The CDF of image tracking error when lamps are occluded.

It can be seen that, for 90% of the data, the tracking error increases significantly as the occluded increases. The tracking error of the two lamps occluded is also larger than that of the single light. But even if the two lights are occluded simultaneously, 90% of the tracking error is less than 5.25cm. These results demonstrate the excellent robustness of our proposed algorithm.

### 3.2.4 Real-time Ability

Real-time ability is also an important indicator for measuring a VLP system, which determines the maximum allowable speed of the positioning terminal. The complexity of the algorithm, the computational speed of the computing hardware, the size of the image and the number of identification lamps are all factors that affect real-time ability. Our proposed algorithm uses the UKF algorithm to predict the most likely location of the target in the current frame based on previous information. Then use the Cam-shift algorithm to start searching for the target from the predicted area. It is especially important in multi-target tracking by reducing the search area to improve the efficiency of search. The general mean shift has a large overhead in computation time, of which time complexity is O(Tn2). n is the number of sample points, and T is the number of iterations. Combined with the UKF, the time complexity of the mean shift search can be effectively reduced by reducing the number of iterations. In our double-lamp tracking experiment, we calculated that the average processing time for 100 frames is 91.63ms per frame. Compared to our previous algorithm, the average processing time for one frame is 0.162 seconds [10].

Table 2. Comparison of time for different algorithms to process one frame.

|  | ellipse fitting | pixel intensity detection | Cam-shift with UKF |
|---|---|---|---|
| Average time to process a frame | 715.24ms | 297.53ms | 91.63ms |

There are many factors that affect the speed of the positioning algorithm, such as video resolution, encoding format, transmission method, processor speed, and so on. The experimental conditions of different papers may be different. Therefore, in order to make the discussion more meaningful, we compared the average time of one frame processing between the ellipse fitting and pixel intensity detection algorithm and our proposed algorithm, as shown in the table. The results show that our proposed algorithm significantly improves the speed of positioning.

# 4 Conclusion

In this paper, we propose a multi-target image tracking algorithm using improved Cam-shift with UKF. The results of the image tracking algorithm are then combined with the positioning algorithm to obtain the position of the terminal. We carefully distinguished the errors caused by the image tracking algorithm and the positioning algorithm. For image tracking algorithm, we use the tracking error defined in the proposition to evaluate. For positioning algorithm, we use the positioning error defined to evaluate. With UKF predicting the most probable position of ROIs through the current motion states, the proposed improved Cam-shift greatly improves the search efficiency of ROI regions, thereby significantly enhancing the real-time performance of the algorithm. The characteristics of the Cam-shift algorithm ensure that even if the LED is partially occluded, it can give relatively correct results robustly. Although our positioning accuracy has not reached the current state of the art, it has reached a satisfactory level for VLP systems. Our proposed method perfectly balances the accuracy, real-time, and robustness.

For the experiment, the proposed VLP method can achieve a tracking accuracy of up to 0.61cm for double-lamp tracking. In dynamic tracking and positioning, our proposed method achieves the proposed positioning error of 3.29cm, which is not much different from the compared algorithm. Moreover, the tracking error when less than 50% and from 50% to 90% of the area of the single lamp is occluded and two lamp occluded simultaneously are 2.41 cm, 4.21 cm and 5.25 cm, respectively. The average processing time per frame is 91.63 ms. The above results show that the VLP method we proposed has high positioning accuracy, good real-time performance and strong robustness. In addition, from the error analysis, the accuracy of the three image tracking algorithms is not significantly different. This suggests that the main error of our VLP system comes from the positioning algorithm. This points the way for future efforts to reduce errors.

## Acknowledgments

This research was funded by Special Funds for the Cultivation of Guangdong College Students' Scientific and Technological Innovation("Climbing Program" Special Funds) grant number pdjh2017b0040, pdjha0028, pdjh2019b0037 and "The One Hundred-Step Ladder Climbing Program" grant number j2tw201902166, j2tw201902189.

**Conflicts of Interest:** The authors declare no conflicts of interest.

# References

1. Kim, N.; Jing, C.; Zhou, B.; Kim, Y. Smart parking information system exploiting visible light communication. Int. J. Smart Home 2014, 8, 251–260.

2. Al Nuaimi, K.; Kamel, H. A survey of indoor positioning systems and algorithms. In Proceedings of the 2011 International Conference on Innovations in Information Technology (IIT), Abu Dhabi, UAE, 25–27 April 2011; pp. 185–190.

3. Liu, H.; Darabi, H.; Banerjee, P.; Liu, J. Survey of Wireless Indoor Positioning Techniques and Systems. Syst. Man Cybern. 2007, 37, 1067–1080.

4. T.-H. Do, M. Yoo, An in-depth survey of visible light communication based positioning systems, Sens. Basel Switzerland 16 (5) (2016) 678–1–678–40.

5. R. Zhang, W.-D. Zhong, Q. Kian, S. Zhang, A single led positioning system based on circle projection, IEEE Photon. J. 9 (4) (2017) 7905209.

6. Kim Jae-Yoon, Yang Se-Hoon, Son Yong-Hwan, Han Sang-Kook. High-resolution indoor positioning using light emitting diode visible light and camera image sensor [J]. IET Optoelectronics, 2016, 10(05):184-192

7. W. Pan, Y. Hou and S. Xiao, "Visible light indoor positioning based on camera with specular reflection cancellation," 2017 Conference on Lasers and Electro-Optics Pacific Rim (CLEO-PR), Singapore, 2017, pp. 1-4.

8. W. Guan, L. Huang, B. Hussain and C. Patrick Yue, "Robust Robotic Localization using Visible Light Positioning and Inertial Fusion," in IEEE Sensors Journal, doi: 10.1109/JSEN.2021.3053342.

9. Kim J , Kang S , Lee S . VLC-based location data transferal for smart devices. Optical Switching and Networking, 2017, 23, 250- 258.

10. Guan, W.; Chen, X.; Huang, M.; Liu, Z.; Wu, Y.; Chen, Y. High Speed Robust Dynamic Positioning and Tracking Method Based on Visual Visible Light Communication Using Optical Flow Detection and Bayesian Forecast. IEEE Photon. J. 2018, 10, 10–1109.

11. Liu, Zhipeng & Guan, Weipeng & Wen, Shangsheng. (2019). Improved Target Signal Source Tracking and Extraction Method Based on Outdoor Visible Light Communication Using an Improved Particle Filter Algorithm Based on Cam-Shift Algorithm. IEEE Photonics Journal. PP. 1-1. 10.1109/JPHOT.2019.2940773.

12. Yan, Z.; Liang, W.; Lv, H. A Target Tracking Algorithm Based on Improved Camshift and UKF. J. Softw. Eng. Appl. 2014, 7, 132014.

13. Peng, Q.; Guan, W.; Wu, Y.; Cai, Y.; Xie, C.; Wang, P. Three-dimensional high-precision indoor positioning strategy using Tabu search based on visible light communication. Opt. Eng. 2018, 57, 016101.

14. Chen, Shihuan & Guan, Weipeng & Tan, Zequn & Wen, Shangsheng & Liu, Manxi & Wang, Jingmin & Li, Jingyi. (2019). High accuracy and error analysis of indoor visible light positioning algorithm based on image sensor.

15. Danakis, Christos & Afgani, Mostafa & Povey, Gordon & Underwood, I. & Haas, Harald. (2012). Using a CMOS camera sensor for visible light communication. 1244-1248. 10.1109/GLOCOMW.2012.6477759.

16. Xie, C.; Guan, W.; Wu, Y.; Fang, L.; Cai, Y. The LED-ID detection and recognition method based on visible light positioning using proximity method. IEEE Photon. J. 2018, 10, 7902116.

17. Fang, J.; Yang, Z.; Long, S.; Wu, Z.; Zhao, X.; Liang, F.; Jiang, Z.L.; Chen, Z. High-speed indoor navigation system based on visible light and mobile phone. IEEE Photon. J. 2017, 9, 8200711.